\documentclass[showpacs]{revtex4} 
\usepackage{amsmath,amssymb,graphicx,textcomp,upgreek} 
\usepackage{natbib}
\usepackage[english]{babel}

\begin{document}
\vspace{0mm}
\title{ DEBYE MODEL FOR THE SURFACE PHONONS } %
\author{Yu.M. Poluektov}
\email{yuripoluektov@kipt.kharkov.ua} %
\affiliation{National Science Center ``Kharkov Institute of Physics
and Technology'', 1, Akademicheskaya St., 61108
Kharkov, Ukraine} %

\begin{abstract}
A quantum description of the surface waves in an isotropic elastic
body without the use of the  semiclassical quantization is proposed.
The problem about the surface waves is formulated in the Lagrangian
and Hamiltonian representations. Within the framework of the
generalized Debye model, the contribution of the surface phonons
(``rayleighons'') to thermodynamic functions is calculated. It is
emphasized that the role of the surface phonons can be significant
and even decisive in low-dimensional systems, granular and porous
media, and that their contribution to the total heat capacity
increases with decreasing temperature.
\newline%
\end{abstract}
\maketitle

\section{Introduction}\vspace{-0mm} 
Methods of describing spatially homogeneous condensed media are
fairly well developed, a large number of specific problems have been
solved. Although here, too, there remain many unsolved problems
associated mainly with the correct account for interparticle
interactions and the description of phase transitions.

In recent years, more and more attention has been paid to both
experimental and theoretical study of media wherein interphase
boundaries and surfaces play an important role. It is precisely in
such essentially inhomogeneous systems that there is reason to hope
for the discovery of new physical effects and the production of new
materials with unique properties on their basis. A simple transfer
of the methods of theoretical study of homogeneous systems to the
case of strongly inhomogeneous media is clearly insufficient, and
therefore there is a need for the elaboration and development of
special approaches. For low temperatures, the quantum-mechanical
description of such systems, in particular of their electronic and
phonon properties, has acquired a fundamental importance.

A widespread model of solids, which gives a correct description of
thermodynamic properties in the limiting cases of low and high
temperatures, is the well-known model proposed by Debye even before
the creation of modern quantum theory [1]. In the Debye model for an
elastic isotropic medium, it is assumed that phonons have a certain
averaged velocity. However, it offers no principal difficulty to
generalize the Debye model for an isotropic solid with taking into
account longitudinal and transverse phonons propagating with
different velocities [2]. As was shown by Rayleigh [3], in an
isotropic elastic body, together with longitudinal and transverse
waves, along the body surface there may propagate surface waves, the
amplitude of which decreases with distance from the boundary. The
quasiparticles corresponding to these waves -- the surface phonons
-- will also give a contribution to the thermodynamic functions of a
solid which is proportional to the area of its surface. This
contribution can be significant and even decisive for
low-dimensional samples, as well as bodies having a granular or
porous structure. The questions of the theory of local and surface
vibrations of a crystal lattice, as well as the literature on this
issue, are considered in the book [4].

The quantization of the surface waves, that is similar to the
semiclassical quantization of the photon field by Planck, was
carried out by Khalatnikov in the study of heat transfer between a
solid and superfluid helium [5]. However, a consistent quantum
theory, as is known, was built only more than a quarter of a century
after the introduction of a new world constant by Planck. At the
same time, a general prescription for the transition from classical
to quantum description became clear. It consists in the fact that
within the framework of the classical approach there are selected
generalized coordinates and canonical momenta, which, upon passing
to the quantum description, are considered as operators acting in
the space of wave functions and obeying the commutation relations
which generalize the classical Poisson brackets. This quantization
prescription is suitable both for systems of particles interacting
by means of a potential [6,7] and for continuous media (fields)
[2,8,9]. For surface waves at the boundary of an isotropic solid,
such a quantization method has not yet been implemented, which is
the main goal of this work. On the basis of the generalized Debye
model, taking into account the difference in the velocities of
longitudinal and transverse sound waves [2], we calculate the
contribution of the surface phonons to thermodynamic functions at
arbitrary temperatures.

\section{Surface waves in an isotropic elastic medium}\vspace{-0mm}
Let us first consider the classical surface wave problem. Suppose
that an elastic isotropic medium fills the half-space $z<0$. Thus,
the surface separating the medium and vacuum is $z=0$. The
deformation vector in an elastic isotropic medium ${\bf u}({\bf
r},z,t)$ satisfies the equation [10]
\begin{equation} \label{01}
\begin{array}{l}
\displaystyle{%
  \ddot{{\bf u}}=c_t^2\Delta{\bf u}\,+\big(c_l^2-c_t^2\big)\nabla\mbox{div}{\bf u}, %
}
\end{array}
\end{equation}
where
\begin{equation} \label{02}
\begin{array}{l}
\displaystyle{%
  c_t=\sqrt{\frac{\mu}{\rho}}, \qquad c_l=\sqrt{\frac{\lambda+2\mu}{\rho}}
}
\end{array}
\end{equation}
are the velocities of the transverse and longitudinal waves, $\mu,
\lambda$ are the Lame coefficients, $\rho$ is the density.

Since the system under consideration is spatially homogeneous in the
$xy$-plane and inhomogeneous along the $z$-axis perpendicular to the
surface, we represent the real deformation vector in the form of an
expansion in two-dimensional plane waves
\begin{equation} \label{03}
\begin{array}{l}
\displaystyle{%
  {\bf u}({\bf r},z,t)=\frac{1}{\sqrt{A}}\sum_{{\bf k}}\Big[ {\bf u}({\bf k},z,t)e^{i{\bf k}{\bf r}}+{\bf u}^*({\bf k},z,t)e^{-i{\bf k}{\bf
  r}}\Big],
}
\end{array}
\end{equation}
$A$ is the surface area. Here ${\bf r}$ and ${\bf k}$  are the
two-dimensional vectors: ${\bf r}\equiv(x,y)$, ${\bf
k}\equiv(k_x,k_y)$. As is known [10], the deformation vector of a
medium can be represented as the sum ${\bf u}={\bf u}_l+{\bf u}_t$,
where ${\rm{div}}{\bf u}_t=0$ and ${\rm{rot}}{\bf u}_l=0$. Each of
these vectors can be expressed in terms of scalar and vector
potentials
\begin{equation} \label{04}
\begin{array}{l}
\displaystyle{%
  {\bf u}_l=\nabla\varphi, \qquad  {\bf u}_t={\rm{rot}}\boldsymbol{\uppsi} 
}
\end{array}
\end{equation}
which satisfy the wave equations
\begin{equation} \label{05}
\begin{array}{l}
\displaystyle{%
  \ddot{\varphi}=c_l^2\Delta\varphi, \qquad \ddot{\boldsymbol{\uppsi}}=c_t^2\Delta\boldsymbol{\uppsi}. %
}
\end{array}
\end{equation}
The potentials can be represented in a form similar to the
decomposition (3):
\begin{equation} \label{06}
\begin{array}{l}
\displaystyle{%
  \varphi({\bf r},z,t)=\frac{1}{\sqrt{A}}\sum_{{\bf k}}\Big[ \varphi({\bf k},z,t)e^{i{\bf k}{\bf r}}+\varphi^*({\bf k},z,t)e^{-i{\bf k}{\bf r}}\Big], %
}\vspace{2mm}\\ %
\displaystyle{%
  \boldsymbol{\uppsi}({\bf r},z,t)=\frac{1}{\sqrt{A}}\sum_{{\bf k}}\Big[ \boldsymbol{\uppsi}({\bf k},z,t)e^{i{\bf k}{\bf r}}+\boldsymbol{\uppsi}^*({\bf k},z,t)e^{-i{\bf k}{\bf r}}\Big]. %
}%
\end{array}
\end{equation}
Applying the Fourier transform in time
\begin{equation} \label{07}
\begin{array}{l}
\displaystyle{%
  \varphi({\bf k},z,t)=\int_{-\infty}^{\infty}\varphi({\bf k},z,\omega)e^{-i{\bf \omega}t}d\omega, \qquad %
 \boldsymbol{\uppsi}({\bf k},z,t)=\int_{-\infty}^{\infty}\boldsymbol{\uppsi}({\bf k},z,\omega)e^{-i{\bf \omega}t}d\omega, %
}
\end{array}
\end{equation}
from the equations (4) we find the solutions bounded at $z<0$
\begin{equation} \label{08}
\begin{array}{l}
\displaystyle{%
  \varphi({\bf k},z,\omega)=A({\bf k},\omega)\exp(\gamma_l z), \qquad %
  \boldsymbol{\uppsi}({\bf k},z,\omega)={\bf B}({\bf k},\omega)\exp(\gamma_t z), %
}
\end{array}
\end{equation}
where
\begin{equation} \label{09}
\begin{array}{l}
\displaystyle{%
  \gamma_l=\sqrt{k^2-\frac{\omega^2}{c_l^2}}, \qquad \gamma_t=\sqrt{k^2-\frac{\omega^2}{c_t^2}}. %
}
\end{array}
\end{equation}
If the radicands in (9) are positive, then the waves decay
exponentially deep into the medium, and if the radicands are
negative, then the solutions (8) have the form of oscillations.

The components of the vector of longitudinal deformations, with
taking into account the form of the solutions (8) and the formula
(4), are determined by the relations:
\begin{equation} \label{10}
\begin{array}{l}
\displaystyle{%
  u_{lx}=\frac{i}{\sqrt{A}}\sum_{{\bf k}}\int\!d\omega k_x \Big[A({\bf k},\omega)e^{\gamma_lz}e^{i({\bf k}{\bf r}-\omega t)}-A^*({\bf k},\omega)e^{\gamma_l^*z}e^{-i({\bf k}{\bf r}-\omega t)}\Big], %
}\vspace{2mm}\\ %
\displaystyle{%
   u_{ly}=\frac{i}{\sqrt{A}}\sum_{{\bf k}}\int\!d\omega k_y \Big[A({\bf k},\omega)e^{\gamma_lz}e^{i({\bf k}{\bf r}-\omega t)}-A^*({\bf k},\omega)e^{\gamma_l^*z}e^{-i({\bf k}{\bf r}-\omega t)}\Big], %
}\vspace{2mm}\\ %
\displaystyle{%
   u_{lz}=\frac{1}{\sqrt{A}}\sum_{{\bf k}}\int\!d\omega \Big[\gamma_lA({\bf k},\omega)e^{\gamma_lz}e^{i({\bf k}{\bf r}-\omega t)}+\gamma_l^*A^*({\bf k},\omega)e^{\gamma_l^*z}e^{-i({\bf k}{\bf r}-\omega t)}\Big]. %
}%
\end{array}
\end{equation}
The components of the vector of transverse deformations, with
account of (8) and (4), are given by the formulas:
\begin{equation} \label{11}
\begin{array}{l}
\displaystyle{%
  u_{tx}=\frac{1}{\sqrt{A}}\sum_{{\bf k}}\int\!d\omega\Big\{\big[ik_yB_z({\bf k},\omega)-\gamma_tB_y({\bf k},\omega)\big]e^{\gamma_tz}e^{i({\bf k}{\bf r}-\omega t)}+\!\big[-ik_yB_z^*({\bf k},\omega)-\gamma_t^*B_y^*({\bf k},\omega)\big]e^{\gamma_t^*z}e^{-i({\bf k}{\bf r}-\omega t)}\Big\}, %
}\vspace{2mm}\\ %
\displaystyle{%
  u_{ty}=\frac{1}{\sqrt{A}}\sum_{{\bf k}}\int\!d\omega\Big\{\big[-ik_xB_z({\bf k},\omega)+\gamma_tB_x({\bf k},\omega)\big]e^{\gamma_tz}e^{i({\bf k}{\bf r}-\omega t)}+\!\big[ik_xB_z^*({\bf k},\omega)+\gamma_t^*B_x^*({\bf k},\omega)\big]e^{\gamma_t^*z}e^{-i({\bf k}{\bf r}-\omega t)}\Big\}, %
}\vspace{2mm}\\ %
\displaystyle{%
  u_{tz}=\frac{i}{\sqrt{A}}\sum_{{\bf k}}\int\!d\omega\Big\{\big[k_xB_y({\bf k},\omega)-k_yB_x({\bf k},\omega)\big]e^{\gamma_tz}e^{i({\bf k}{\bf r}-\omega t)}-\!\big[k_xB_y^*({\bf k},\omega)-k_yB_x^*({\bf k},\omega)\big]e^{\gamma_t^*z}e^{-i({\bf k}{\bf r}-\omega t)}\Big\}. %
}
\end{array}
\end{equation}
The formulas (10), (11) give a general solution for the deformation
vector under arbitrary boundary conditions, which is bounded in the
half-space $z<0$.

Let us obtain a particular solution, provided that no external
forces act on the surface. The external force acting per unit
surface area is determined by the expression $f_i=\sigma_{ik}n_k$
[10], where ${\bf n}$ is the outward normal vector to the surface,
which in the case under consideration is directed along the
$z$-axis, so that ${\bf n}=(0,0,1)$. Assuming that at $z=0$ the
force is equal to zero, we have the boundary conditions
$\sigma_{xz}=\sigma_{yz}=\sigma_{zz}=0$, which, with taking into
account the form of the stress tensor for an isotropic elastic
medium, have the form
\begin{equation} \label{12}
\begin{array}{l}
\displaystyle{%
  \frac{\partial u_x}{\partial z}+\frac{\partial u_z}{\partial x}=0, \qquad %
  \frac{\partial u_y}{\partial z}+\frac{\partial u_z}{\partial y}=0, \qquad %
  (1-\sigma)\frac{\partial u_z}{\partial z}+\sigma\!\left(\frac{\partial u_x}{\partial x}+\frac{\partial u_y}{\partial y}\right)=0, %
}
\end{array}
\end{equation}
where $\sigma$ is the Poisson coefficient. Using the formulas (10),
(11), by means of the boundary conditions (12) we obtain the
algebraic equations for the expansion coefficients in these formulas
\begin{equation} \label{13}
\begin{array}{l}
\displaystyle{%
  2ik_x\gamma_lA + k_xk_yB_x-\!\big(\gamma_t^2+k_x^2\big)B_y+ik_y\gamma_tB_z=0, %
}\vspace{2mm}\\ %
\displaystyle{%
  2ik_y\gamma_lA +\!\big(\gamma_t^2+k_y^2\big)B_x-k_xk_yB_y-ik_x\gamma_tB_z=0, %
}\vspace{2mm}\\ %
\displaystyle{%
  \left[(1-\sigma)\gamma_l^2-\sigma\big(k_x^2+k_y^2\big)\right]\!A + i(1-2\sigma)\gamma_t\big(k_xB_y-k_yB_x\big)=0.  %
}%
\end{array}
\end{equation}
To shorten the notation we used the designations $A({\bf
k},\omega)\equiv A, {\bf B}({\bf k},\omega)\equiv {\bf B}$. It is
also convenient to introduce the designations
\begin{equation} \label{14}
\begin{array}{l}
\displaystyle{%
  B_+=k^{-1}\big(k_xB_x+k_yB_y\big), \qquad B_-=k^{-1}\big(k_xB_y-k_yB_x\big), %
}
\end{array}
\end{equation}
where $k^2=k_x^2+k_y^2$. Then the system of equations (13) takes the
form
\begin{equation} \label{15}
\begin{array}{l}
\displaystyle{%
  iB_z=\alpha_tB_+, %
}%
\end{array}
\end{equation}
\vspace{-7mm}
\begin{equation} \label{16}
\begin{array}{l}
\displaystyle{%
  2i\alpha_lA-\!\big(1+\alpha_t^2\big)B_-=0, %
}\vspace{2mm}\\ %
\displaystyle{%
  (2-y)A+2i\alpha_tB_-=0, %
}
\end{array}
\end{equation}
wherein
\begin{equation} \label{17}
\begin{array}{l}
\displaystyle{%
  y\equiv\frac{\omega^2}{c_t^2k^2},\qquad \gamma_l=\alpha_l k, \qquad \gamma_t=\alpha_t k, %
}\vspace{2mm}\\ %
\displaystyle{%
  \alpha_l\equiv\sqrt{1-\xi^2y}, \qquad \alpha_t\equiv\sqrt{1-y}, \qquad \xi^2=\left(\frac{c_t}{c_l}\right)^{\!2}=\frac{(1/2-\sigma)}{(1-\sigma)}. %
}
\end{array}
\end{equation}
The compatibility condition for a system of linear homogeneous
equations gives the equation
\begin{equation} \label{18}
\begin{array}{l}
\displaystyle{%
  4\alpha_t\alpha_l=(2-y)^2, %
}%
\end{array}
\end{equation}
which, with account of the formulas (17), is reduced to the
well-known cubic equation [10]
\begin{equation} \label{19}
\begin{array}{l}
\displaystyle{%
  y^3-8y^2+8(3-2\xi^2)y-16(1-\xi^2)=0, %
}%
\end{array}
\end{equation}
determining the dispersion law of surface waves. The parameter $y$
is a real number less than one, the value of which is determined by
the ratio of  the transverse and longitudinal wave velocities. Thus,
the velocity of surface waves
\begin{equation} \label{20}
\begin{array}{l}
\displaystyle{%
  c_s=\sqrt{y}\,c_t %
}%
\end{array}
\end{equation}
is less than the velocity of body waves. The radicands in (17) are
positive for all values of $y$ and the quantities
$\alpha_l,\alpha_t$ are real, while $\alpha_l>\alpha_t$. In the case
when there is a relationship $\omega=\omega_0({\bf k})$ between the
frequency and the wavenumber, the amplitudes in the formulas (10),
(11) should be taken in the form
\begin{equation} \label{21}
\begin{array}{l}
\displaystyle{%
  A({\bf k},\omega)=\tilde{A}({\bf k})\delta\big(\omega-\omega_0({\bf k})), \qquad %
  {\bf B}({\bf k},\omega)=\tilde{{\bf B}}({\bf k})\delta\big(\omega-\omega_0({\bf k})), %
}%
\end{array}
\end{equation}
where $\tilde{A}({\bf k})\equiv A\big({\bf k},\omega_0({\bf
k})\big), \,\tilde{{\bf B}}({\bf k})\equiv {\bf B}\big({\bf
k},\omega_0({\bf k})\big)$. In our case $\omega_0({\bf
k})=\sqrt{y}c_t k$. In the following we will omit the tilde sign.
The relations (13)\,--\,(16) make it possible to express the
deformation vector only in terms of one amplitude $A({\bf k})$:
\begin{equation} \label{22}
\begin{array}{l}
\displaystyle{%
  u_{x}({\bf r},z,t)=\frac{i}{\sqrt{A}}\sum_{{\bf k}} \Big\{ k_x g_1(z,k) \Big[A({\bf k})e^{i({\bf k}{\bf r}-\omega_0({\bf k})t)}-A^*({\bf k})e^{-i({\bf k}{\bf r}-\omega_0({\bf k})t)}\Big]\Big\}, %
}\vspace{2mm}\\ %
\displaystyle{%
 u_{y}({\bf r},z,t)=\frac{i}{\sqrt{A}}\sum_{{\bf k}} \Big\{ k_y g_1(z,k) \Big[A({\bf k})e^{i({\bf k}{\bf r}-\omega_0({\bf k})t)}-A^*({\bf k})e^{-i({\bf k}{\bf r}-\omega_0({\bf k})t)}\Big]\Big\},  %
}\vspace{2mm}\\ %
\displaystyle{%
 u_{z}({\bf r},z,t)=\frac{\alpha_l}{\sqrt{A}}\sum_{{\bf k}} \Big\{ k\,g_2(z,k) \Big[A({\bf k})e^{i({\bf k}{\bf r}-\omega_0({\bf k})t)}+A^*({\bf k})e^{-i({\bf k}{\bf r}-\omega_0({\bf k})t)}\Big]\Big\}. %
}%
\end{array}
\end{equation}
Here, the following functions are defined
\begin{equation} \label{23}
\begin{array}{l}
\displaystyle{%
  g_1(z,k)\equiv e^{\alpha_lkz}-\frac{2\alpha_l\alpha_t}{1+\alpha_t^2}e^{\alpha_tkz}, \qquad %
  g_2(z,k)\equiv e^{\alpha_lkz}-\frac{2}{1+\alpha_t^2}e^{\alpha_tkz}. %
}%
\end{array}
\end{equation}
After defining the time-dependent amplitudes
\begin{equation} \label{24}
\begin{array}{l}
\displaystyle{%
  A({\bf k},t)=A({\bf k})e^{-i\omega_0({\bf k})t}, \qquad A^*({\bf k},t)=A^*({\bf k})e^{i\omega_0({\bf k})t}, %
}%
\end{array}
\end{equation}
and introducing the new amplitudes
\begin{equation} \label{25}
\begin{array}{l}
\displaystyle{%
  X({\bf k},t)=X^*(-{\bf k},t)\equiv A({\bf k},t)+A^*(-{\bf k},t), %
}%
\end{array}
\end{equation}
the components of the deformation vector can be written as
\begin{equation} \label{26}
\begin{array}{l}
\displaystyle{%
  u_{x}({\bf r},z,t)=\frac{i}{\sqrt{A}}\sum_{{\bf k}} k_x\,g_1(z,k)X({\bf k},t)e^{i{\bf k}{\bf r}}, %
}\vspace{2mm}\\ %
\displaystyle{%
  u_{y}({\bf r},z,t)=\frac{i}{\sqrt{A}}\sum_{{\bf k}} k_y\,g_1(z,k)X({\bf k},t)e^{i{\bf k}{\bf r}}, %
}\vspace{2mm}\\ %
\displaystyle{%
  u_{z}({\bf r},z,t)=\frac{\alpha_l}{\sqrt{A}}\sum_{{\bf k}} k\,g_2(z,k)X({\bf k},t)e^{i{\bf k}{\bf r}}. %
}%
\end{array}
\end{equation}
Before proceeding to quantizing surface excitations, we present the
Lagrangian and the Hamiltonian formulations of the problem.

\section{Lagrangian and Hamiltonian formulations of the surface waves problem} %
The density of the Lagrangian function, which leads to the equations
for waves in an isotropic elastic medium (1), has the form
\begin{equation} \label{27}
\begin{array}{l}
\displaystyle{%
  \Lambda({\bf r},z,t)=\frac{\rho}{2}\,\dot{{\bf u}}^2-\frac{\lambda}{2}\big(\mbox{div}{\bf u}\big)^2-\frac{\mu}{2}\big(\nabla_j u_i \nabla_i u_j + \nabla_j u_i \nabla_j u_i\big), %
}
\end{array}
\end{equation}
and to find the complete Lagrangian function one should integrate
(27) over the entire volume occupied by the medium
\begin{equation} \label{28}
\begin{array}{l}
\displaystyle{%
  L=\int_{A}d{\bf r}\int_{-\infty}^{0}dz\Lambda({\bf r},z). %
}
\end{array}
\end{equation}
Using the obtained expressions for the components of the deformation
vector (26), for which the boundary conditions (12) are fulfilled,
and integrating over spatial coordinates, we obtain the Lagrangian
function in the form
\begin{equation} \label{29}
\begin{array}{l}
\displaystyle{%
  L=\frac{\rho\alpha_l}{4}\Theta(y)\sum_{{\bf k}}\!\Big[k \big|\dot{X}({\bf k})\big|^2 - c_s^2k^3\big|X({\bf k})\big|^2 \Big], %
}
\end{array}
\end{equation}
where
\begin{equation} \label{30}
\begin{array}{l}
\displaystyle{%
  \Theta(y)\equiv\frac{\left[16(1-y)^2+(2-y)^4\right]}{(1-y)(2-y)^4}-2\frac{(2+y)}{(2-y)}=\frac{y^2\big(8-16y+11y^2-2y^3\big)}{(1-y)(2-y)^4}. %
}
\end{array}
\end{equation}\newpage\noindent
Instead of the complex quantities in (29), one should pass to the
real quantities $X({\bf k})=X'({\bf k})+iX''({\bf k})$ and consider
$X'({\bf k}), X''({\bf k})$ as generalized coordinates and
$\dot{X}'({\bf k}), \dot{X}''({\bf k})$ as generalized velocities,
on which the Lagrangian function depends. However, it should be
taken into account that the Lagrangian must be expressed in terms of
independent coordinates and velocities. Meanwhile, due to the
condition (25), the coordinates at the oppositely directed wave
vectors ${\bf k}$ and $-{\bf k}$ are linked by the relations
\begin{equation} \label{31}
\begin{array}{l}
\displaystyle{%
  X'({\bf k},t)=X'(-{\bf k},t), \qquad X''({\bf k},t)=-X''(-{\bf k},t) %
}%
\end{array}
\end{equation}
and therefore are not independent. To pass to independent variables
in the Lagrangian (29), we decompose it into two terms, each of
which contains a summation over all wave vectors except for the
oppositely directed ones. Due to the conditions (31), the
contribution to the Lagrangian of each such term will be the same.
Therefore, the Lagrangian (29) should be multiplied by two, and the
summation should be carried out only over those wave vectors among
which there are no the oppositely directed ones. For example, we may
sum over all wave vectors having $k_z>0$. As a result, the
Lagrangian, expressed in terms of the real independent even and odd
(31) coordinates and velocities, takes the form
\begin{equation} \label{32}
\begin{array}{l}
\displaystyle{%
  L=\frac{\rho\alpha_l\Theta(y)}{2}\sum_{{\bf k}}^{\frown}\Big[k \dot{X}'^{2}({\bf k}) - c_s^2k^3X'^{2}({\bf k}) \Big]+ %
  \frac{\rho\alpha_l\Theta(y)}{2}\sum_{{\bf k}}^{\frown}\Big[k \dot{X}''^{2}({\bf k}) - c_s^2k^3X''^{2}({\bf k}) \Big]. %
}
\end{array}
\end{equation}
Here, the symbol $\frown$ above the sum sign, as explained above,
means a summation over vectors among which there are no the
oppositely directed ones. The Euler-Lagrange equations give rise to
the equations of motion
\begin{equation} \label{33}
\begin{array}{l}
\displaystyle{%
  \ddot{X}'({\bf k})+c_s^2k^2X'({\bf k})=0, \qquad \ddot{X}''({\bf k})+c_s^2k^2X''({\bf k})=0. %
}%
\end{array}
\end{equation}

Let us move on to the Hamiltonian description, defining the
canonical momenta
\begin{equation} \label{34}
\begin{array}{l}
\displaystyle{%
  \Pi'({\bf k})=\frac{\partial L}{\partial\dot{X}'({\bf k})}=\rho\alpha_l\Theta(y)k\dot{X}'({\bf k}), \qquad  %
  \Pi''({\bf k})=\frac{\partial L}{\partial\dot{X}''({\bf k})}=\rho\alpha_l\Theta(y)k\dot{X}''({\bf k}).   %

}%
\end{array}
\end{equation}
We introduce the Hamilton function
\begin{equation} \label{35}
\begin{array}{l}
\displaystyle{%
  H=\sum_{{\bf k}}^{\frown}\left(\dot{X}'({\bf k})\frac{\partial L}{\partial\dot{X}'({\bf k})} %
  + \dot{X}''({\bf k})\frac{\partial L}{\partial\dot{X}''({\bf k})}\right) - L = H'+H'', %
}%
\end{array}
\end{equation}
where $H', H''$ are the Hamiltonians expressed in terms of the
coordinates and momenta with one and two primes:
\begin{equation} \label{36}
\begin{array}{l}
\displaystyle{%
  H'=\frac{\rho\alpha_l\Theta(y)}{2}\sum_{{\bf k}}^{\frown}\left[\frac{\Pi'^2(\bf k)}{\rho^2\alpha_l^2\Theta^2(y)}+c_s^2k^3X'^2(\bf k) \right], %
}\vspace{2mm}\\ %
\displaystyle{%
  H''=\frac{\rho\alpha_l\Theta(y)}{2}\sum_{{\bf k}}^{\frown}\left[\frac{\Pi''^2(\bf k)}{\rho^2\alpha_l^2\Theta^2(y)}+c_s^2k^3X''^2(\bf k) \right]. %
}%
\end{array}
\end{equation}
Since these Hamiltonians have the same form, it suffices to consider
only the Hamiltonian expressed in terms of the coordinates and
momenta with one prime. From the Hamilton equations
\begin{equation} \label{37}
\begin{array}{l}
\displaystyle{%
  \dot{\Pi}'({\bf k})=-\frac{\partial H'}{\partial X'({\bf k})}, \qquad  %
  \dot{X}'({\bf k})=\frac{\partial H'}{\partial\Pi'({\bf k})}   %
}%
\end{array}
\end{equation}
there follow the equations of motion
\begin{equation} \label{38}
\begin{array}{l}
\displaystyle{%
  \dot{\Pi}'({\bf k})=-\rho\alpha_l\Theta(y)c_s^2k^3X'({\bf k}), \qquad  %
  \dot{X}'({\bf k})=\frac{\Pi'(\bf k)}{\rho\alpha_l\Theta(y)k}.   %
}%
\end{array}
\end{equation}
These equations with account of the definition of momenta (34), of
course, can also be written in the form of the second-order
equations for an oscillator (33).

\section{Quantization of surface waves} %
Now we pass from the classical to the consistently quantum
description of surface excitations without the use of the
semiclassical approximation [5]. In this case, the coordinates and
momenta should be considered as operators obeying the following
well-known commutation relations:
\begin{equation} \label{39}
\begin{array}{l}
\displaystyle{%
  \big[X'({\bf k}),\Pi'({\bf k}')\big]\equiv X'({\bf k})\Pi'({\bf k}')-\Pi'({\bf k}')X'({\bf k})=i\hbar\Delta({\bf k}-{\bf k}'),  %
}\vspace{2mm}\\ %
\displaystyle{%
  \big[X'({\bf k}),X'({\bf k}')\big]=\big[\Pi'({\bf k}),\Pi'({\bf k}')\big]=0.  %
}%
\end{array}
\end{equation}
The same commutation relations hold for the coordinates and momenta
with two primes, and all variables with one prime commute with all
variables with two primes. It is convenient to pass from the
Hermitian operators $X'({\bf k})=X'^+({\bf k})$ and $\Pi'({\bf
k})=\Pi'^+({\bf k})$ to the new non-Hermitian operators $a^+({\bf
k}), a({\bf k})$, which, as we will see, have the meaning of the
operators of creation and annihilation of surface phonons
\begin{equation} \label{40}
\begin{array}{l}
\displaystyle{%
  X'({\bf k})=C({\bf k})a({\bf k})+C^*({\bf k})a^+({\bf k}),  %
}\vspace{2mm}\\ %
\displaystyle{%
  \Pi'({\bf k})=B({\bf k})a({\bf k})+B^*({\bf k})a^+({\bf k}),  %
}%
\end{array}
\end{equation}
where the coefficients $C({\bf k}), B({\bf k})$ are $c$\,-numbers.
The commutation relations (39) will be satisfied if the following
well-known commutation relations are required to hold
\begin{equation} \label{41}
\begin{array}{l}
\displaystyle{%
  \big[a({\bf k}),a^+({\bf k}')\big]\equiv a({\bf k})a^+({\bf k}')-a^+({\bf k}')a({\bf k})=\Delta({\bf k}-{\bf k}'),  %
}\vspace{2mm}\\ %
\displaystyle{%
  \big[a({\bf k}),a({\bf k}')\big]=\big[a^+({\bf k}),a^+({\bf k}')\big]=0.  %
}%
\end{array}
\end{equation}
In addition, in order for the transformation (40) to be canonical,
the following condition should hold
\begin{equation} \label{42}
\begin{array}{l}
\displaystyle{%
  C({\bf k})B^*({\bf k})-C^*({\bf k})B({\bf k})=i\hbar.  %
}%
\end{array}
\end{equation}
In the Hamilton operators (36), which we write in the form
(similarly for $H''$)
\begin{equation} \label{43}
\begin{array}{l}
\displaystyle{%
  H'=\sum_{{\bf k}}^{\frown}\Big[\psi(k)\Pi'^2({\bf k})+\varphi(k)X'^2({\bf k}) \Big], %
}%
\end{array}
\end{equation}
where for brevity the designations $\psi(k)\equiv
1\big/2\rho\alpha_l\Theta(y)k$, $\varphi(k)\equiv
\rho\alpha_l\Theta(y)c_s^2k^3\big/2$ are used, it is also necessary
to pass to the new operators $a^+({\bf k}), a({\bf k})$. When
substituting the relations (40) into (43) we require that only the
operators of the form $a({\bf k})a^+({\bf k})$ or $a^+({\bf
k})a({\bf k})$ remain in the resulting Hamiltonian, while the
operators $a^2({\bf k})$ and $a^{+2}({\bf k})$  drop out. This
entails the fulfillment of the condition
\begin{equation} \label{44}
\begin{array}{l}
\displaystyle{%
  \psi(k)B^2({\bf k})+\varphi(k)C^2({\bf k})=0.  %
}%
\end{array}
\end{equation}
The condition (44) will be satisfied if
\begin{equation} \label{45}
\begin{array}{l}
\displaystyle{%
  B^2(k)=2\big/\hbar\rho\alpha_l\Theta(y)c_sk^2, \qquad   %
  C^2(k)=-\hbar\big/2\rho\alpha_l\Theta(y)c_sk^2.   %

}%
\end{array}
\end{equation}
When extracting the root the signs can be chosen arbitrarily, since
this will not affect the final result. Thus, the canonical
transformation (40) takes the form
\begin{equation} \label{46}
\begin{array}{l}
\displaystyle{%
  X'({\bf k})=i\frac{\hbar}{k\sqrt{\hbar\rho\alpha_l 2\Theta(y)c_s}}\big[a({\bf k})-a^+({\bf k})\big],  %
}\vspace{2mm}\\ %
\displaystyle{%
  \Pi'({\bf k})= k\sqrt{\frac{\hbar\rho\alpha_l\Theta(y)c_s}{2}}\big[a({\bf k})+a^+({\bf k})\big]. %
}%
\end{array}
\end{equation}
As a result, using the commutation relations (41), we find
\begin{equation} \label{47}
\begin{array}{l}
\displaystyle{%
  H'=\sum_{{\bf k}}^{\frown} \hbar c_sk\!\left[a^+({\bf k})a({\bf k})+\frac{1}{2}\right]. %
}%
\end{array}
\end{equation}
The diagonalization of the Hamiltonian $H''$ can be carried out in a
similar way. When calculating the thermodynamic quantities, the
Hamiltonians $H'$ and $H''$ make the same contribution, so that one
can use the doubled operator (47) and the total Hamiltonian takes
the form
\begin{equation} \label{48}
\begin{array}{l}
\displaystyle{%
  H=2\sum_{{\bf k}}^{\frown} \hbar c_sk\!\left[a^+({\bf k})a({\bf k})+\frac{1}{2}\right]. %
}%
\end{array}
\end{equation}
The Hamiltonian (48) differs from the Hamiltonian of bulk phonons in
the value of the velocity and in that the summation is carried out
over two-dimensional wave vectors. The surface two-dimensional
phonons, having velocity $c_s$, can naturally be called
``rayleighons''. Note that the quanta of vibrations of the surface
of liquid helium are called riplons [11,12].

\section{Contribution of ``rayleighons'' to thermodynamic functions} %
The average values of the surface thermodynamic functions are
calculated by means of the statistical operator
\begin{equation} \label{49}
\begin{array}{l}
\displaystyle{%
  \rho=\exp\beta\big(F-H\big),   %
}%
\end{array}
\end{equation}
where the Hamiltonian is defined by the formula (48), and
$\beta=1\big/T$ is the inverse temperature. From the normalization
condition $\mbox{Sp}\,\rho=1$ we find the surface free energy
\begin{equation} \label{50}
\begin{array}{l}
\displaystyle{
  F=\frac{\hbar}{2}\sum_{{\bf k}} c_sk\, + T\sum_{{\bf k}}\ln\!\big(1-e^{-\beta\hbar c_sk}\big).  %
}%
\end{array}
\end{equation}
Here the summation, as noted above, is carried out over the
two-dimensional wave vectors ${\bf k}\equiv(k_x,k_y)$. The first
term in (50) determines the contribution of zero oscillations. In
(50) we pass from summation to integration in which the upper limit
of integration over the magnitude of the wave vector is determined,
as in the bulk case, by the Debye relation
$k_D=\big(6\pi^2n\big)^{1/3},\,n=N/V$ is the particle number
density. As a result, we obtain the surface free energy in the form
\begin{equation} \label{51}
\begin{array}{l}
\displaystyle{
  F=\frac{Ak_D^2}{12\pi}\left\{\Theta_s+\frac{3}{2}\,T\!\left[2\ln\!\left(1-e^{-\tau^{-1}}\right)-D_2\!\left(\tau^{-1}\right)\right]\right\},  %
}%
\end{array}
\end{equation}
where $\Theta_s\equiv\hbar c_sk_D$   is the ``surface'' Debye
energy, $\tau\equiv T/\Theta_s$, and the Debye functions are defined
by the formula
\begin{equation} \label{52}
\begin{array}{l}
\displaystyle{
  D_n(x)=\frac{n}{x^n}\int_0^x\frac{z^ndz}{e^z-1} \qquad (n\geq 1).  %
}%
\end{array}
\end{equation}
The surface entropy $S=-\big(\partial F/\partial T\big)_{V\!A}$ has
the form
\begin{equation} \label{53}
\begin{array}{l}
\displaystyle{
  S=\frac{Ak_D^2}{8\pi}\left[3D_2\!\left(\tau^{-1}\right) - 2\ln\!\left(1-e^{-\tau^{-1}}\right)\right].  %
}%
\end{array}
\end{equation}
From here we find the heat capacity at a constant volume and surface
area
\begin{equation} \label{54}
\begin{array}{l}
\displaystyle{
  C_{V\!A}=T\!\left(\frac{\partial S}{\partial T}\right)_{\!V\!A}=\frac{Ak_D^2}{4\pi\tau}\!\left[3\tau D_2\!\left(\tau^{-1}\right) - \frac{2}{e^{\tau^{-1}}-1}\right].  %
}%
\end{array}
\end{equation}

Let us consider the behavior of the surface heat capacity (54) in
the limit of low and high temperatures. Since $D_2(x)\approx
4x^{-2}\zeta(3)$\, at $x\ll 1$, we find at $\tau\ll 1$:
\begin{equation} \label{55}
\begin{array}{l}
\displaystyle{
  C_{V\!A}\approx\frac{3\zeta(3)}{\pi}Ak_D^2\!\left(\frac{T}{\Theta_s}\right)^{\!2}=\frac{3\zeta(3)}{\pi\hbar^2c_s^2}AT^2,  %
}%
\end{array}
\end{equation}
$\zeta(s)$ is the Riemann zeta function. Note that the structure of
this formula is similar to the structure of the corresponding
formula for the bulk case of the usual Debye theory
\begin{equation} \label{56}
\begin{array}{l}
\displaystyle{
  C_{V}\approx\frac{2\pi^2}{5}Vk_D^3\!\left(\frac{T}{\Theta_D}\right)^{\!3}=\frac{2\pi^2}{5\hbar^3c_D^3}VT^3,  %
}%
\end{array}
\end{equation}
where $c_D$ is the average phonon velocity according to Debye
[13,14]. In the two-parameter theory [2], the formula for the
low-temperature bulk heat capacity takes the form
\begin{equation} \label{57}
\begin{array}{l}
\displaystyle{
  C_{V}\approx\frac{2\pi^2}{5}Vk_D^3\cdot f(\chi)\!\left(\frac{T}{\Theta}\right)^{\!3}=\frac{2\pi^2}{5\hbar^3}\left(\frac{3}{2c_t^2+c_l^2}\right)^{\!3/2}Vf(\chi)\,T^3,  %
}%
\end{array}
\end{equation}
where $\,\displaystyle{\Theta^2}=\frac{1}{3}\big(2\Theta_t^2+\Theta_l^2\big), \,\mbox{tg}\,\chi=\sqrt{2}\,\frac{\Theta_t}{\Theta_l}, \,\Theta_t=\hbar c_tk_D, \,\Theta_l=\hbar c_lk_D, \,f(\chi)\equiv\frac{1}{3^{5/2}}\left(\frac{1}{\cos^3\chi}+\frac{2^{5/2}}{\sin^3\chi}\right).$ %

The calculation of the surface contribution to the phonon heat
capacity has received a considerable attention. The quadratic
dependence on temperature was obtained in [15], and subsequently the
form of the coefficient before $T^2$ was refined in other works [4].
In [16\,--\,18],  there was obtained a formula for the surface heat
capacity in various approaches, which in the notation of (17), (19),
(20) can be represented in the form [6,16]
\begin{equation} \label{58}
\begin{array}{l}
\displaystyle{
  \overline{C}_{A}=\frac{3\zeta(3)}{\pi\hbar^2c_s^2}\cdot y\frac{\big(2\xi^4-3\xi^2+3\big)}{4\big(1-\xi^2\big)}\,AT^2.  %
}%
\end{array}
\end{equation}
In the formulas (55), (58)  the coefficients  at $T^2$ differ in
value insignificantly. So, for the Poisson coefficient $\sigma=1/2$
and $\xi^2=0$ we have $\overline{C}_{A}\big/C_{V\!A}\approx 0.685$,
and for $\sigma=0$ and $\xi^2=1/2$ the ratio
$\overline{C}_{A}\big/C_{V\!A}\approx 0.763$. The calculation of the
surface heat capacity according to the formula (55) gives a slightly
higher value of the heat capacity than the calculation according to
the formula (58). Despite the small quantitative difference, the
question about the reasons for the difference between the early
formula (58) and the seemingly more natural formula (55) requires an
additional consideration.

In the case of high temperatures $\tau>1$ the surface heat capacity,
like the bulk heat capacity, tends to a constant value:
\begin{equation} \label{59}
\begin{array}{l}
\displaystyle{
  C_{V\!A}\approx\frac{Ak_D^2}{4\pi}\left[1-\frac{1}{24}\left(\frac{\Theta_s}{T}\right)^{\!2}\right].  %
}%
\end{array}
\end{equation}
Let there be a sample in the shape of a cube with the edge length
$L$, volume $V=L^3$ and surface area $A\approx L^2$. Let us estimate
the ratio of the surface and bulk heat capacities in such a sample.
At high temperatures, when $C_V=3N$, we have
\begin{equation} \label{60}
\begin{array}{l}
\displaystyle{
  \frac{C_{V\!A}}{C_V}\approx\frac{Ak_D^2}{4\pi}\frac{1}{3N}=a\frac{l}{L}, %
}%
\end{array}
\end{equation}
where $l=n^{-1/3}$ is the average distance between atoms,
$a=\big(6\pi^2\big)^{1/3}\big/12\pi\approx 0.4$. Thus, at high
temperatures for macroscopic samples $L\gg l$ the surface heat
capacity is small in comparison with the bulk heat capacity.

Let us consider the case of low temperatures, when for the bulk heat
capacity the relation
$\displaystyle{C_V\approx\frac{12\pi^4}{5}N\!\left(\frac{T}{\Theta}\right)^{\!3}}$ %
holds.  In this case
\begin{equation} \label{61}
\begin{array}{l}
\displaystyle{
  \frac{C_{V\!A}}{C_V}\approx\frac{15\zeta(3)}{\pi^4}\frac{Ak_D^2}{12\pi N}\left(\frac{\Theta}{\Theta_s}\right)^{\!2}\frac{\Theta}{T}. %
}%
\end{array}
\end{equation}
With decreasing temperature this ratio increases, and at a certain
temperature $T_*$ the surface and bulk heat capacities become equal.
From (61) it follows that
\begin{equation} \label{62}
\begin{array}{l}
\displaystyle{
  \frac{T_*}{\Theta}=\frac{15\zeta(3)}{\pi^4}\,a\frac{l}{L}\left(\frac{\Theta}{\Theta_s}\right)^{\!2}\approx 0.075\frac{l}{L}, %
}%
\end{array}
\end{equation}
where it is assumed that $\Theta\approx\Theta_s$. For samples with
sizes of the order of a nanometer $L\approx 10^{-7}\,\mbox{\it cm}$,
and, taking into account that $l\approx 10^{-8}\,\mbox{\it cm}$, the
ratio $T_*\big/\Theta\approx 0.01$. Since the Debye energy has the
order of magnitude $\Theta\approx 10^{2}\,\mbox{\it K}$, then the
law for the heat capacity $ C_{V\!A}\approx T^2$ in samples of such
size can be observed at helium temperatures.  The contribution of
the surface heat capacity can also be the main one in porous and
granular materials.

In the bulk case the low-temperature heat capacity, according to the
Debye theory [1], is proportional to the cube of temperature, and in
the two-dimensional case, when the contribution of the surface
phonons is taken into account, it is proportional to the square of
temperature. This indicates that the temperature exponent in the
low-temperature behavior of the heat capacity coincides with the
dimension of space. In this regard, it is interesting to calculate
the entropy and heat capacity of ``one-dimensional'' phonons. Such a
situation with the quasi-one-dimensional phonons could be realized
in long cylindrical filaments of small radius. The propagation of
the classical surface waves under conditions of cylindrical geometry
was studied in many works [19]. For a qualitative study of a
one-dimensional system of phonons, it is sufficient to assume that
in the expression for the free energy (50) the summation is
performed over wave vectors oriented in one direction. In this case,
we find
\begin{equation} \label{63}
\begin{array}{l}
\displaystyle{
  F=\frac{Lk_D}{4\pi}\left\{\Theta_1+4T\!\left[\ln\!\left(1-e^{-\tau^{-1}}\right)-D_1\!\left(\tau^{-1}\right)\right]\right\},  %
}%
\end{array}
\end{equation}
where $L$ is the length of a sample, $\Theta_1=\hbar c_1k_D$, $c_1$
is the velocity of the ``one-dimensional'' phonons, and as before
$k_D=\big(6\pi^2n\big)^{1/3}$. From (63) there follow the
expressions for the ``one-dimensional'' entropy and heat capacity:
\begin{equation} \label{64}
\begin{array}{l}
\displaystyle{
  S=\frac{Lk_D}{\pi}\left[2D_1\!\left(\tau^{-1}\right) - \ln\!\left(1-e^{-\tau^{-1}}\right)\right],  %
}%
\end{array}
\end{equation}
\vspace{-5mm}
\begin{equation} \label{65}
\begin{array}{l}
\displaystyle{
  C_{V\!L}=T\!\left(\frac{\partial S}{\partial T}\right)_{\!V\!L}=\frac{Lk_D}{\pi\tau}\!\left[2\tau D_1\!\left(\tau^{-1}\right) - \frac{1}{e^{\tau^{-1}}-1}\right].  %
}%
\end{array}
\end{equation}
In the low-temperature limit $\tau\ll 1$, taking into account that
$D_1(x)\approx\pi^2\big/6x$, from here we get:
\begin{equation} \label{66}
\begin{array}{l}
\displaystyle{
  S=C_{V\!L}\approx\frac{\pi L k_D}{3}\!\left(\frac{T}{\Theta_1}\right).  %
}%
\end{array}
\end{equation}
As could be expected, the low-temperature entropy and heat capacity
of a system of ``one-dimensional'' phonons are proportional to
temperature.

\section{Conclusion}\vspace{-0mm} %
A method for quantizing surface elastic waves in an isotropic solid
without the use of the semiclassical approximation is proposed, and
in the Debye approach the contribution of the surface phonons
(``rayleighons'') to thermodynamic functions is calculated. In
agreement with the previous works [15\,--\,18] it is shown that in
the limit of low temperatures the contribution of the surface
phonons is proportional to the square of temperature, while the
value of the proportionality coefficient is somewhat different from
the earlier results [15\,--\,18]. It is also shown that in the
one-dimensional case at low temperatures the dependence of the heat
capacity is linear. Thus, the exponent in the temperature dependence
of the phonon heat capacity in the low-temperature limit is
determined by the spatial dimensionality of a system.

There are two extreme points of view on the Debye model. Often this
model is given an unduly fundamental meaning and, when processing
experimental data, observable quantities are adjusted to the
relations of theory assuming that the Debye energy depends on
temperature. The opposite point of view is that the relations of the
Debye theory are considered as rough interpolation formulas [13,14].
The Debye model, of course, is an approximate and rather simple
(which is its value) model of the solid body, but there is reason to
assert that its value is not limited only to the possibility of
constructing a single interpolation formula that would correctly
describe the behavior of the solid body in the limit of low and high
temperatures. This model allows further development and
generalization, for example, accounting for the difference in the
velocities of longitudinal and transverse phonons [2], the
interaction of phonons [8,9], and, as shown in this article, it can
be extended to describe surface phenomena in solids.

When analyzing thermodynamic and kinetic properties of crystals
whose anisotropy is not large and the considered effects are not
associated with the existence of singled-out directions in crystals,
it is possible to use with a good accuracy a more simple model of an
isotropic medium after  choosing its parameters in an optimal way
[20]. It was shown in [2] that the previously proposed method of
describing the elastic properties of crystals on the basis of a
comparison with an isotropic medium [20] follows from the
requirement of the maximal closeness of the free energies of a
crystal and an isotropic medium. The two-parameter Debye model for
an isotropic medium with effective elastic moduli [2] can be a good
approximation for describing the properties of crystals. All the
general remarks made above refer, in particular, to the further
development of the considered in this article Debye model for the
surface phonons.

The author is grateful to A.S. Kovalev for helpful comments.


\newpage
\begin{thebibliography}{99}
\bibitem{Debye}
P.\,Debye, Ann. Phys. \textbf{39}(4), 789 (1912).  
$\mbox{doi:10.1002/andp.19123441404}$
\bibitem{Poluektov1}
Yu.M.\,Poluektov, East Eur. J. Phys. \textbf{5}, \textnumero 3, 4
(2018); arXiv:2004.06658v1\,[cond-mat.stat-mech].
\bibitem{Rayleigh}
J.\,Rayleigh, Proc. London Math. Soc. \textbf{s1-17}(1), 4 (1885).
$\mbox{doi:10.1112/plms/s1-17.1.4}$
\bibitem{Maradudin}
A.\,Maradudin, Defects and vibrational spectrum of crystals, Mir,
Moscow, 432\,p. (1968).
\bibitem{Khalatnikov}
I.M.\,Khalatnikov, Theory of superfluidity, Nauka, Moscow, 320\,p. (1971). %
\bibitem{MM}
A.\,Maradudin, E.\,Montroll, G.\,Weiss, Theory of lattice dynamics
in the harmonic approximation, Acad.\,Press, 319\,p. (1968).
\bibitem{Kosevich}
A.M.\,Kosevich, Foundation of crystal lattice mechanics, Nauka,
Moscow, 280\,p. (1972). %
\bibitem{Poluektov2}
Yu.M.\,Poluektov, Low Temp. Phys. \textbf{41}, 922 (2015).
$\mbox{doi:10.1063/1.4936228}$
\bibitem{Poluektov3}
Yu.M.\,Poluektov, East Eur. J. Phys. \textbf{3}, \textnumero 3, 35 (2016). %
\bibitem{LL1}
L.D.\,Landau, E.M.\,Lifshitz, Theory of elasticity, Vol.\,7,
Butterworth-Heinemann (3rd ed.), 196\,p. (1986).
\bibitem{Shikin}
V.B.\,Shikin, Y.P.\,Monarkha, Two-dimensional charged systems in
helium, Nauka, Moscow, 156\,p. (1989).
\bibitem{Monarkha}
Y.\,Monarkha, K.\,Kono, Two-dimensional Coulomb liquids and solids,
Springer-Verlag, New York, 350\,p. (2004).
\bibitem{LL2}
L.D.\,Landau, E.M.\,Lifshitz, Statistical Physics, Vol.\,5 (Part 1),
Butterworth-Heinemann (3rd ed.), 544\,p. (1980).
\bibitem{AM}
N.\,Ashcroft, N.\,Mermin, Solid state physics, Harcourt College
Publishers, 826\,p. (1976).
\bibitem{Brager}
A.\,Brager, A.\,Schuchowitzky, Journ. Chem. Phys. \textbf{14}, 569 (1946). %
$\mbox{doi:10.1063/1.1724202}$
\bibitem{Dupuis}
M.\,Dupuis, R.\,Mazo, L.\,Onzager, Journ. Chem. Phys. \textbf{33}, 1452 (1960). %
$\mbox{doi:10.1063/1.1731426}$
\bibitem{Stratton}
R.\,Stratton, Journ. Chem. Phys. \textbf{37}, 2972 (1962).
$\mbox{doi:10.1063/1.1733127}$
\bibitem{MW}
A.A.\,Maradudin, R.F.\,Wallis, Phys. Rev. \textbf{148}, 945 (1966).
$\mbox{doi:10.1103/PhysRev.148.945}$
\bibitem{Viktorov}
I.A.\,Viktorov, Sound surface waves in solids, Nauka, Moscow, 287\,p. (1981). %
\bibitem{Fedorov}
F.I.\,Fedorov, Theory of elastic waves in crystals, Springer, New
York, 375\,p. (1968).

\end{thebibliography}
\end{document}